# The Role of Global Value Chains in Carbon Intensity Convergence: A Spatial Econometrics Approach

*May 22, 2021*

**Kazem Biabany Khameneh**

Faculty of Management and Economics, Tarbiat Modares University
Email: kazem.biabany@modares.ac.ir
Tel: +98(21)82884640
Address: Nasr, Jalal AleAhmad, Tehran, Iran

**Reza Najarzadeh**

Faculty of Management and Economics, Tarbiat Modares University
Email: najarzar@modares.ac.ir
Tel: +98(21)82884640
Address: Nasr, Jalal AleAhmad, Tehran, Iran
P.O. Box: 14115-111
*Corresponding Author*

**Hassan Dargahi**

Faculty of Economics and Political Sciences, Shahid Beheshti University
Email: h-dargahi@sbu.ac.ir
Tel: +98(21)29902977
Address: Evin, Tehran, Iran
P.O. Box: 193 581 3654

**Lotfali Agheli**

Economic Research Institute, Tarbiat Modares University
Email: aghelik@modares.ac.ir
Tel: +98(912)6186317
Address: Nasr, Jalal AleAhmad, Tehran, Iran
P.O. Box: 14115-111

# The Role of Global Value Chains in Carbon Intensity Convergence: A Spatial Econometrics Approach

**Abstract:** The expansion of trade agreements has provided a potential basis for trade integration and economic convergence of different countries. Moreover, developing and expanding global value chains (GVCs) have provided more opportunities for knowledge and technology spillovers and the potential convergence of production techniques. This can result in conceivable environmental outcomes in developed and developing countries. This study investigates whether GVCs can become a basis for the carbon intensity (CI) convergence of different countries. To answer this question, data from 101 countries from 1997 to 2014 are analyzed using spatial panel data econometrics. The results indicate a spatial correlation between GVCs trade partners in terms of CI growth, and they confirm the GVCs-based conditional CI convergence of the countries. Moreover, estimates indicate that expanding GVCs even stimulates bridging the CI gap between countries, *i.e.*, directly and indirectly through spillover effects. According to the results, GVCs have the potential capacity to improve the effectiveness of carbon efficiency policies. Therefore, different dimensions of GVCs and their benefits should be taken into account when devising environmental policies.

**Keywords:** global value chains, carbon intensity, conditional convergence, spatial panel data regression, trade in value-added

# 1. Introduction

The impacts of trade on the environment and greenhouse gas emissions have been among the controversial topics in the economic literature over the past three decades, and it is still difficult to determine whether trade is good for the environment or not (Frankel and Rose, 2005). Apart from how trade affects the environment, trade agreements and integrations have raised the likelihood for the potential convergence of environmental indicators of various countries (Apergis and Payne, 2020; Baghdadi et al., 2013). Meanwhile, dramatic changes in trade and globalization processes, such as the emergence of GVCs and the international fragmentation of production, have provided more opportunities for trade integration and further expansion of knowledge and technology flows among countries.

This study aims to determine if GVCs can provide a basis for the CI convergence of countries and how expanding participation in GVCs affects CI growth. The answers to these questions can somewhat clarify how this new form of trade can bridge the CI gap between countries, and how expanding participation in GVCs can be a strategy to reduce environmental degradation. The answers might even reveal that this new form of trade is not only detrimental, but also beneficial to the environment. As acknowledged by Thomakos and Alexopoulos (2016), CI is the strongest explanatory variable in the Environmental Performance Index[1] (EPI). In other words, these two are two sides of the same coin. Therefore, CI changes can properly measure the environmental performance of countries. If we find out how GVCs affect the CI changes of countries, we can then appropriately generalize it to environmental performance, benefit from GVC capacities to formulate environmental protection policies and regulations, and reach multilateral agreements to mitigate climate change.

---

[1] https://epi.yale.edu/

The rest of this paper is organized as follows: Sections 2 and 3 review the theoretical foundations and the research background. The methodology for the data analysis is introduced in Section 4, and the results are presented in Section 5. Finally, Section 6 discusses the results and draws the conclusions.

## 2. Theoretical Foundations

There are different views on the link between trade and the environment in the economic literature. Some refer to the positive environmental impacts of trade through technology development and transfer and spillover effects, while others discuss the pollution haven hypothesis (PHH) and the potential degradation of the environment in developing countries through trade with developed countries. However, some argue that trade is detrimental to the environment because it scales up the economic activities that are inherently harmful to the environment and encourages the polluting countries to engage more extensively in highly-polluting activities (Copeland and Taylor, 2013; Weber et al., 2021). Thus, trade is a double-edged sword that can both help protect the environment through the development and spillover of eco-friendly technologies (Lovely and Popp, 2011; Nemati et al., 2019) and be a ground for environmental degradation, as suggested by the pollution haven hypothesis and the factor endowment theory (Antweiler et al., 2001; Cherniwchan et al., 2017; Shen, 2008).

The recent increase in the number of trade agreements and the emergence of GVCs, which account for more than half of the global trade, have fundamentally altered the structure and organization of international trade, encouraging the establishment of international production networks and massive developments in bilateral trade, especially for intermediate goods. This new wave of globalization has also redrawn the boundaries of knowledge, production structure, and comparative advantages of different countries, while changing the relations between developed and developing countries (Baldwin, 2017). GVCs have facilitated the transfer of technical innovations, skills, and knowledge from developed countries to developing countries (Jangam and Rath, 2020), allowing developing countries to enter global markets and reap potential benefits without having to develop ancillary industries and resorting to the local production of all the required inputs (Rodrik, 2018).

Countries have already been participating in international trade (*e.g.*, developing countries import parts and technology to produce and supply goods to the domestic market). However, the new form of trade has increased the countries' use of international production networks, intensified the knowledge information flows (Taglioni et al., 2016), and facilitated further knowledge spillovers compared to the traditional trade methods (Piermartini and Rubínová, 2014). Knowledge flows more easily within the supply chain because the outsourcer transfers the necessary knowledge and technology to the input producer firm to ensure the efficient production of inputs and compliance with its production standards (Piermartini and Rubínová, 2014).

By reallocating scarce resources to the most profitable activities, GVCs can increase the efficiency of an economy by improving static efficiency (*i.e.*, modifying existing processes and capacities) and dynamic efficiency (*i.e.*, creating new processes and capacities). As shown by Kummritz (2016), the expansion of GVCs is expected to raise per capita income, investment, productivity, and domestic value-added production. Participation in supply chains and international production networks can also lead to learning-by-doing benefits, economies of scale, and technology progress and spillovers, and even accelerate the industrialization process and the development of the country's service sector (Bernhardt and Pollak, 2016; Dorrucci et al., 2019; Ignatenko et al., 2019; Kummritz, 2016; Pigato et al., 2020; Taglioni et al., 2016). Markusen (1984) also argues that the increasing global technical efficiency is associated with the growth of multinational enterprises (a key element of GVCs) as there is evidence of technology spillovers following their activities (Keller, 2010).

The outcomes and effects of expanding participation in GVCs are beyond economic growth. For example, the growth of GVCs and trade in intermediate inputs can improve South-South trade (Hanson, 2012). Moreover, expanding GVCs can intensify the transmission of shocks, the synchronization of global business cycles, and changes in specialization patterns, while also causing transformations in international trade policies (Dorrucci et al., 2019; Wang et al., 2017). However, the effects of GVCs on the environmental performance of countries and their role in environmental phenomena have received little attention in the relevant literature.

If GVCs facilitate the convergence of production techniques as expected (Rodrik, 2018), integration of GVCs with technological advancement and transfer will increase the countries' income level, thus possibly helping to converge some of their economic indicators (Ignatenko et al., 2019). The existing evidence also suggests that trade, trade integration, and regional cooperation help increase the convergence of energy efficiency, energy intensity, and environmental performance of the countries (Han et al., 2018; Qi et al., 2019; Wan et al., 2015). Meanwhile, trade integration requires changes in the existing standards and regulations (even environmental standards and regulations) of the countries to reduce trade frictions and facilitate trade flows (Nicoletti et al., 2003). Holzinger *et al*. (2008) also showed that increasing international links would lead to the convergence of the involved countries' environmental policies. The expansion of multinational corporations also contributes to sustainable management, knowledge transfer, and the development of low-CI production technologies (López et al., 2019).

On the other hand, the new form of trade may not only help countries converge by equalizing the prices of production factors, but also improve the environmental performance of the countries by increasing technology spillovers and environmentally-efficient knowledge. Along with technical progress, the spillover, diffusion, and transfer of cleaner technologies to countries with lower energy and environmental efficiency, especially developing countries, will further improve environmental performance (Gerlagh and Kuik, 2014; Huang et al., 2020; Jaffe et al., 2002; LeSage and Fischer, 2008; Wan et al., 2015). Jiang *et al*. (2019) reported that countries with technologically-advantageous trading partners emitted less carbon because they were allowed to share the produced resources with their major trading partners. Additionally, the flow of knowledge between firms in GVCs can accelerate advances in eco-friendly technologies. The expansion of GVCs can also bring more affordable and less expensive technologies for the generation and consumption of clean and more efficient sources of energy (WDR, 2020).

Therefore, expanding GVCs can potentially provide more opportunities for the convergence of the environmental performance of the involved countries. Nevertheless, confirming or rejecting such hypotheses will require empirical tests, which is the purpose of this study.

## 3. Research Background

To the best of our knowledge, the role of GVCs in converging the environmental performance of countries has not yet been studied. However, some studies have dealt with the effects of trade and foreign direct investment (FDI) on the convergence of energy intensity and carbon emissions. For example, Wan *et al*. (2015) analyzed the effects of trade spillovers on the energy efficiency convergence of 16 EU countries using spatial panel data. Their results indicated a correlation (spatial autocorrelation) between two trading partners in terms of energy efficiency. They confirmed the existence of a convergence and found that countries with greater reliance on trade had higher energy efficiency growth and a higher convergence rate because trade facilitated competition and technological influences, preventing the occurrence of the race to the bottom.

Jiang *et al*. (2018) studied the convergence of energy intensity in Chinese provinces in the period from 2003 to 2015 using a spatial regression model. They concluded that neglecting spatial spillovers would lead to the underestimation of conditional convergence. They argued that reduced energy intensity depended on the internal factors of each province, while it was also influenced by spatial spillover effects of the neighbors, especially technology and knowledge spillovers and input-output relationships. Their findings also revealed that FDI played a major role in reducing energy intensity.

You and Lv (2018) studied the spatial effects of economic globalization on $CO_2$ emissions in 83 countries over the period from 1985 to 2013 using a spatial panel approach. They found a spatial correlation in the $CO_2$ emission of countries. Moreover, they concluded that having highly-globalized neighbors would improve the environmental quality of a country, indicating that international cooperation and economic integration are important for the environment.

Huang *et al*. (2019) conducted a study on 61 countries based on data from 1992 to 2016 using the panel smooth transition regression (PSTR) model. They report that the globalization threshold (KOF index) affects the conditional convergence of energy intensity in a nonlinear manner, *i.e.*, as the globalizations crosses the threshold, convergence speeds up. They suggest that increasing interactions between countries through trade and political, social, and interregional cooperation can reduce energy intensity.

Xin-gang *et al*. (2019) studied the role of FDI in reducing the energy efficiency gap between different regions of China in the period from 2005 to 2014 using spatial econometric models. Their results showed that there was no absolute convergence; however, they found conditional convergence in the energy intensity of the studied regions, which was strengthened by FDI. They also reported the spatial spillover effects of FDI.

Huang *et al*. (2019) analyzed the CI convergence of Chinese provinces in the period from 2000 to 2016 using a dynamic spatial regression approach. They concluded that there was a CI convergence in Chinese provinces, stating that FDI, trade openness, and spatial spillover effects could help reduce energy intensity.

Awad (2019) estimated panel cointegration regressions to determine the effects of intercontinental trade (trade with countries outside Africa) of 46 African countries on $CO_2$ and PM10 emissions from 1990 to 2017. The results indicated that intercontinental trade improved Africa's environmental quality because countries of this continent were able to achieve sustainable development by overcoming challenges such as poverty and internal strife through trade mechanisms.

Qi *et al*. (2019) studied the energy intensity convergence in 56 countries of the Belt and Road Initiative (BRI) using the smooth transition regression (STR) model with an emphasis on the role of China's trade with these countries in energy intensity convergence. The results indicated that trade facilitated energy intensity convergence and the convergence rate was higher in countries with large bilateral trade with or considerable technology imports from China. They also provided evidence of spillover effects of technology.

Some studies have recently analyzed the effects of GVCs on the environment and energy. For instance, Kaltenegger *et al*. (2017) studied the effects of GVCs on the energy footprint in 40 countries using structural decomposition analysis. The results showed that there was an increase of about 29% in the global energy footprint from 1995 to 2009 mainly due to the growth of economic activities. On the other hand, the reduced sectoral energy intensity was the main factor slowing down energy consumption. Changes in GVCs were responsible for a 7.5% increase in the global energy consumption, 5.5% of which was related to the increased backward linkages, and 1.8% could be attributed to changes in the regional composition of intermediate inputs. Although the global economic boom in East Asia has increased the global energy footprint, the sectoral composition of GVCs has negligible effects on the energy footprint.

Liu *et al*. (2018) studied the effects of the GVC position of 14 Chinese manufacturing industries on their energy and environmental efficiency in the period from 1995 to 2009. They concluded that there was a positive feedback relationship between these industries, and the improved GVC position of the industries caused a 35% improvement in their energy and environmental efficiency. They suggested that the allocation of more budget to research and development activities and the promotion of knowledge absorption capacities were necessary to achieve a better GVC position and improve the environment.

Sun *et al*. (2019) analyzed the effects of the GVC position on the carbon efficiency of 60 countries in the period from 2000 to 2011 using data envelopment analysis (DEA) and panel remission. They concluded that the promoted GVC position improved both energy efficiency and carbon efficiency. They also showed that the effects of promoting the GVC position were greater on technological industries than labor-intensive and resource-intensive industries.

Yasmeen *et al*. (2019) studied the effects of trade in value-added (TiVA) on air pollutants in 39 countries in the period from 1995 to 2009. The panel regression results indicated that there was an inverted-U relationship between TiVA and the concentration of air pollutants, and the effects of TiVA were greater on carbon monoxide than other pollutants. They concluded that the early stages of TiVA might increase environmental pollution; however, further development of TiVA and production methods could reduce the concentration of air pollutants.

## 4. Methodology

This study first analyzes whether there is a GVC-based correlation between countries in terms of CI growth. Then, it investigates the role of expanding participation in GVCs in the CI convergence of the countries. The possible spillover effects of participation in GVCs are also addressed. Therefore, it is necessary to employ appropriate statistical techniques that can both consider spatial dimensions of the statistical data and model spillover effects. The spatial statistics used in this study for data analysis will be discussed in the next subsections.

### 4.1. Spatial Autocorrelation

Spatial autocorrelation is based on the assumption that adjacent subjects are more similar to each other. Suppose that countries i and j have a very high volume/value of bilateral TiVA and considerable trade relations. In other words, they are trade neighbors. Can this neighborhood play a role in their correlation in terms of CI growth? Measures such as Moran's I and Geary's C are usually used to examine such a spatial autocorrelation. These two measures show whether there is a correlation between the two countries in terms of CI growth concerning GVCs.

Moran's I for a dataset is calculated using the following equation (Gangodagamage et al., 2008; Kalkhan, 2011, chap. 3):

$$I = \frac{\frac{1}{W}\sum_i \sum_{i \neq j} w_{ij}(z_i - \bar{z}) \cdot (z_j - \bar{z})}{\frac{1}{n}\sum_i (z_i - \bar{z})^2}. \tag{1}$$

In which, $I$ represents Moran's I, $z_i$ denotes CI growth, and $i$ and $j$ represent 1 through n (i.e., the number of sample countries). Moreover, $w_{ij}$ is a spatial weight matrix that shows how close $i$ and $j$ are (this was measured based on GVCs in this study, and the definition of this matrix is presented in Section 4.3), and $W$ is their sum. Similar to Pearson's correlation coefficient, positive and negative values of Moran's I indicate positive and negative spatial autocorrelation, respectively.

A difference between Geary's C and Moran's I is that the former measures the interaction between $i$ and $j$ based on their difference, and not based on standard deviation (Gangodagamage et al., 2008; Kalkhan, 2011, chap. 3):

$$C = \frac{(n-1)\sum_i \sum_j w_{ij}(z_i - z_j)^2}{2 \sum_i \sum_j w_{ij}(z_i - \bar{z})^2}. \tag{2}$$

In this equation, $C$ represents Geary's C, which ranges from 0 to 2. Values smaller than 1 correspond to positive autocorrelation, while values greater than 1 correspond to negative autocorrelation. The significance of both Geary's C and Moran's I can be examined by assuming the spatial randomness under the standard normal distribution.[2]

## 4.2. Spatial Panel Data Regression

Spatial autocorrelation measures the possible correlation of CI growth between countries $i$ and $j$ due to their GVC-based neighborhood. Spatial panel data regression examines whether there is a CI growth convergence and how expanding GVCs affects this convergence. It also identifies the potential spillover effects of GVCs. These models are of special importance as Anselin (2003a) argues that some phenomena, such as neighborhood effects and the race to the bottom, are among the instances of interactions between economic agents,

---

[2] The statistics of Moran's I and Geary's C are $\frac{I-E(I)}{Sd(I)}$ and $\frac{C-E(C)}{Sd(C)}$, respectively, in which, E () and SD denote the expectation and standard deviation, respectively.

suggesting that interactive and spatial models must be used to analyze such situations.

In general, there are three types of commonly-used spatial econometric data panel models, *i.e.*, the Spatial Lag Model (SLM), the Spatial Error Model (SEM), and the Spatial Durbin Model (SDM) (LeSage and Pace, 2009). The SLM model assumes that the observed value of the dependent variable in a section is affected by both exogenous regressors and the spatial mean weight of the dependent variables of its neighbors. The regression can be presented as follows:

$$y_{it} = \rho \sum_{j=1}^{N} w_{ij} y_{jt} + x_{it}\beta + \mu_i + \varepsilon_{it}. \tag{3}$$

In this equation, $y$ represents the dependent variable for section $i$ at time $t$, $\sum_{j=1}^{N} w_{ij} y_{jt}$ denotes endogenous interactive effects of the dependent variables $y_{it}$ and $y_{jt}$ in neighboring sections (N sections), and $\rho$ is the spatial autoregression coefficient that measures the size of the simultaneous spatial correlation between a section and its neighboring sections. Moreover, $x_{it}$ is the matrix of independent explaining variables, and the $\beta$ vector represents the effects of these independent variables (also known as the exogenous regressor), while $\mu_i$ shows the fixed effects that capture the specific effects of each section. The $\varepsilon_{it}$ error component is also assumed to be i.i.d. with a zero mean and a constant variance. In addition, $w_{ij}$ is the spatial weights matrix, which can be established based on both economic and geographical measures (as described in Section 4.3).

SEM also takes into account the interactive effects between the error components. This model is of greater importance when independent variables excluded from the regression affect the interactive effects of the sections. This model can be presented as follows:

$$y_{it} = x_{it}\beta + \mu_i + \lambda \sum_{j=1}^{N} w_{ij}\varphi_{jt} + \varepsilon_{it}. \tag{4}$$

In which, $\varphi_{jt}$ is a component of the spatial autocorrelation error and $\lambda$ denotes the spatial autocorrelation coefficient, which measures the effects of adjacent sections' residuals on the residual of each section.

SDM is a merger of the SEM and SLM models. In this model, $\gamma$ represents the vector of spatial autocorrelation coefficients of explanatory variables. If $\gamma = 0$ and $\rho \neq 0$, the model will be an SAR, and if $\gamma = -\beta\rho$, the model turns into an SEM. Therefore, the SDM model is a testable general specification that includes the other two models, *i.e.*, SEM and SLM (LeSage and Pace, 2009).

Furthermore, the coefficients of the explanatory variables do not accurately reflect the final effects of the variables due to the spatial correlation in spatial regressions. In addition, the model involves both direct and indirect (spillover) effects because of spatial connections and contemporaneous feedback. Hence, it is necessary to estimate the direct and indirect effects, instead of interpreting the coefficients of point estimates. Based on SDM, we have:

$$E(Y_t) = (I_n - \rho W)^{-1}\mu + (I_n - \rho W)^{-1}(X_t\beta + W X_t\gamma). \tag{6}$$

In which, $I_n$ shows the unit n × n matrix, and the spatial multiplier matrix $(I_n - \rho W)^{-1}$ is equal to:

$$(I_n - \rho W)^{-1} = I_n + \rho W + \rho^2 W^2 + \rho^3 W^3 + \cdots \tag{7}$$

Therefore, the matrix of partial derivatives of the dependent variable in different sections with respect to the explanatory variable k in other sections at any time *t* is equal to:

$$\left[\frac{\partial E(Y)}{\partial x_{1k}} \cdots \frac{\partial E(Y)}{\partial x_{Nk}}\right] = \begin{bmatrix} \frac{\partial E(y_1)}{\partial x_{1k}} & \cdots & \frac{\partial E(y_1)}{\partial x_{Nk}} \\ \vdots & \ddots & \vdots \\ \frac{\partial E(y_N)}{\partial x_{1k}} & \cdots & \frac{\partial E(y_N)}{\partial x_{Nk}} \end{bmatrix} \qquad (8)$$

$$= (I_n - \rho W)^{-1} \begin{bmatrix} \beta_k & w_{12}\gamma_k & \cdots & w_{1n}\gamma_k \\ w_{21}\gamma_k & \beta_k & \cdots & w_{2n}\gamma_k \\ \vdots & \vdots & \ddots & \vdots \\ w_{n1}\gamma_k & w_{n2}\gamma_k & \cdots & \beta_k \end{bmatrix}.$$

It can also be rewritten in shorter terms as follows:

$$\frac{\partial E(Y)}{\partial x_k} = (I_n - \rho W)^{-1} S. \qquad (9)$$

Therefore, the mean direct effects of a unit of change in the explanatory variables $x_k$ on $Y$ are obtained through the mean diagonal elements of Matrix S, the mean total effects can be calculated by averaging the sum of the rows or columns of Matrix S, and the mean indirect effects (spillover effects) are equal to the difference between the direct effects and the total effects. Formally, we have:

Total effect:
$$\frac{1}{n}\sum_{i,j}^{n}\frac{\partial E(y_i)}{\partial x_{kj}} = \frac{1}{n}I'_n[(I_n - \rho W)^{-1}S]I_n \qquad (10)$$

Direct effect:
$$\frac{1}{n}\sum_{i}^{n}\frac{\partial E(y_i)}{\partial x_{ki}} = \frac{1}{n}trace[(I_n - \rho W)^{-1}I_n\beta] \qquad (11)$$

Indirect effect:
$$\left(\frac{1}{n}I'_n[(I_n - \rho W)^{-1}S]I_n\right) - \left(\frac{1}{n}trace[(I_n - \rho W)^{-1}I_n\beta]\right) \qquad (12)$$

Accordingly, by estimating the following SDM, the effects of participation in GVCs and the role of GVCs in the CI convergence can be evaluated as follows:

$$\ln \frac{CI_{it}}{CI_{it-1}} = \beta \ln CI_{it-1} + \rho \sum_{j=1}^{N} w_{ij} \ln \frac{CI_{jt}}{CI_{jt-1}} + \theta \ln GVC_{it-1} \qquad (14)$$

$$+ \pi \ln x_{it-1} + \sum_{j=1}^{N} w_{ij} \ln GVC_{it-1} \delta + \mu_i + \varepsilon_{it}.$$

In which, $CI, GVC$, and $x$ denote carbon intensity, GVCs participation, and a vector of other variables describing CI, such as real per capita income (Y), urbanization rate (UR), and energy intensity (EI), respectively (other components are the same as Equation 3). After estimating the above regression using the maximum likelihood (ML) method, the direct, indirect (spillover), and total effects are calculated for each variable to interpret the regression results. Moreover, if the total effect of $\ln CI_{it-1}$ is denoted by B, then B should range from 0 to -1, and it must be statistically significant to establish conditional convergence. The conditional convergence rate can also be obtained using $-\ln(B + 1)$ (LeSage and Fischer, 2008).

### 4.3. Data

This study was conducted on the economic and GVC-based trade data of a set of selected countries. Carbon intensity is calculated by dividing carbon dioxide emissions (kg) by Gross Domestic Product (constant 2010 USD) per year. Moreover, real per capita income (*i.e.*, the gross domestic product at constant 2010 USD divided by population), EI (*i.e.*, energy consumption per kilogram equivalent of crude oil divided by GDP at constant 2010 USD), and urbanization rate (*i.e.*, urban population divided by total population) are added to the model as regressors. The data for 101 countries[3] in the period from 1997 to 2014 were extracted from the World Bank's WDI Database.[4]

Following Aslam *et al*. (2017), the method proposed by Koopman *et al.* (2014) was employed to measure participation in GVCs. In this method, the level of participation of country *i* in GVCs at time *t* is equal to:

$$GVC_{it} = \frac{DVX_{it} + FVA_{it}}{Gross\ Exports_{it}}. \qquad (14)$$

In which, *FVA*, *DVX*, and *Gross Exports* represent foreign value-added, indirect domestic value-added, and the total value of a country's exports in each period, respectively[5]. This indicator measures the level of a country's participation in GVCs regardless of its scale of economy. The data on FVA, DVX, and Gross Exports were extracted from the UNCTAD-EORA Database (Casella et al., 2019)[6].

According to Section 3.1, a spatial weight matrix should be developed to capture the spatial distance between the countries (*i.e.*, sections) in the sample. This weight matrix, which is also called the distance matrix, can be developed either geographically to show the geographical distance between the countries or economically to reveal the economic distance between them (Anselin, 2003b). Similar to studies conducted by Conley *et al*. (2002), Wan *et al*. (2015), Jiang *et*

---

[3] The countries are listed in Appendix A.

[4] The main reason for choosing this number of countries and this time period was the availability of data because spatial data panel regressions should not contain missing data. In other words, only a balanced data panel should be used that includes the same data for all the countries in a given period.

[5] Foreign value-added corresponds to the value added of inputs that were imported in order to produce intermediate or final goods/services to be exported. Indirect domestic value-added is part of the domestic value-added exported to other countries (exported input) that is used as input in other countries' exports. Gross exports or total exports is the sum of GVCs-based trade and traditional trade, which is the same as the export statistics reported by the countries' customs.

[6] Levin et al. (2002) unit root test has been performed, indicating that all variables follow an I(0) process (see Appendix B)

*al.* (2018), and Servén and Abate (2020), the economic distance is employed in this study to develop the spatial weight matrix. Servén and Abate (2020) developed economic distance matrices for pairs of countries based on the size of their bilateral trade. However, the spatial weight matrix in this study will be constructed based on the bilateral TiVA between countries within GVCs[7]:

$$w_{ij} = \frac{Exports_{ij} + Imports_{ij}}{\sum_{K=1}^{K=N} Exports_{ik} + \sum_{K=1}^{K=N} Imports_{ik}} = w_{ji}. \quad (15)$$

In which, *Exports*, *Imports*, and *N* denote the exports of value added from country *i* to country *j*, the imports of value added from country *j* to country *i*, and the total number of countries, respectively. The higher the *w* value between the two countries, the shorter the distance between them and the greater their proximity. Since the sample consists of 101 countries, the spatial weight matrix has 101 rows and columns, and the sum of each row is equal to $1^8$. It should be noted that since the spatial weight matrix is time-invariant, the weight matrix is developed based on the sum of value-added imports and exports of countries during the period from 1997 to 2014. The data on TiVA were extracted from the UNCTAD-EORA Database (Casella et al., 2019), which are calculated using World Input-Output Tables. Figure 1 shows the graph of connectivity for the countries according to Equation 15, where the graph nodes are the countries and the edge between countries i and j is the trade flow between them with a strength of $w_{ij} = w_{ji}$ (the strength of an edge is indicated by its thickness in the graph). Moreover, the size of a node indicates the weighted degree[9] of that node. In addition, countries with a higher $w_{ij} = w_{ji}$ are closest to each other in the graph.

---

[7] The difference between bilateral trade and bilateral TiVA precisely lies in the concept of GVCs, i.e., bilateral trade includes traditional trade and GVCs-based trade, whereas TiVA refers to GVCs-based trade.

[8] This is called row standardization.

[9] The weighted degree is a centrality measurement in graph theory, which measures the number of edges (links) of a node, and the edges are also weighted by their strength.

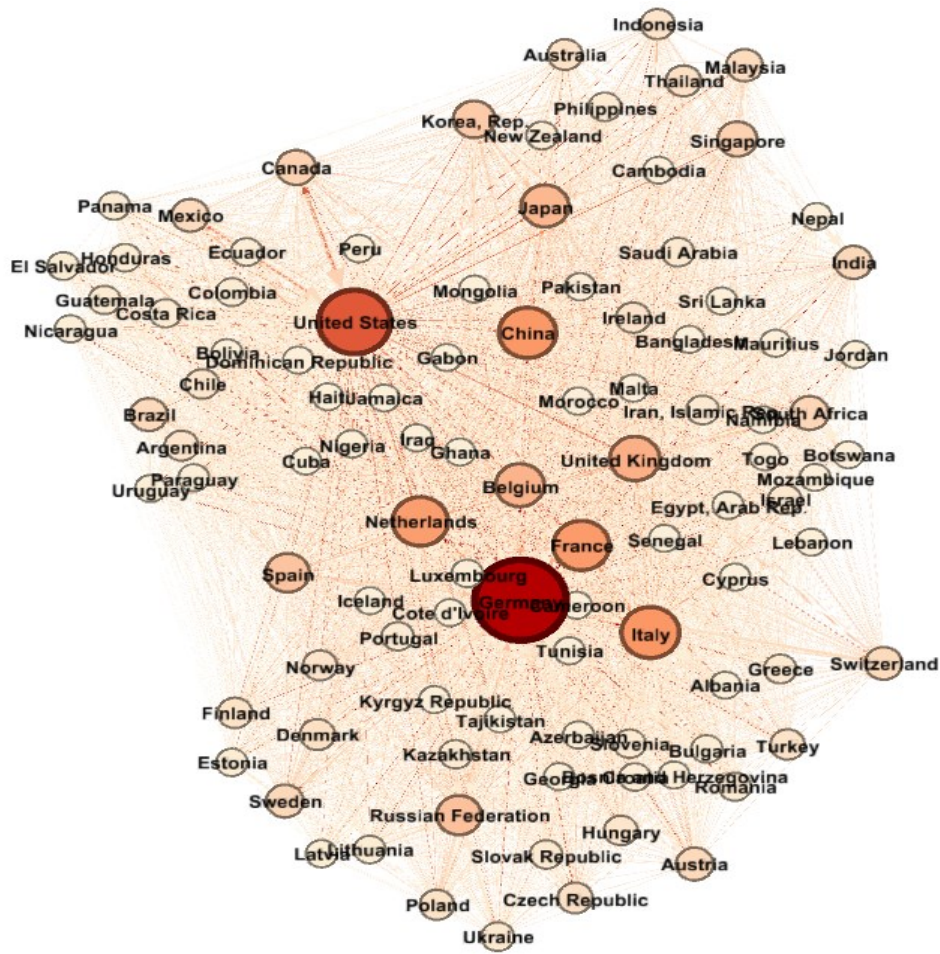

**Figure 1. The connections between countries according to the spatial weight matrix**

5. Results

First, Moran's I and Geary's C are estimated to analyze the GVC-based spatial autocorrelation of the CI growth of the countries. The CI growth of each country from 1997 to 2014 is the variable whose spatial autocorrelation is estimated, and the weight matrix is developed according to Equation 15. The main objective of the study is to determine whether there is a correlation in the CI growth between the countries with bilateral TiVA. Table 1 presents the results of Moran's I and Geary's C. Both coefficients revealed the GVC-based positive spatial autocorrelation of CI growth as the null hypothesis indicating spatial randomness for both tests is rejected at a confidence level of 95%.

As a result, countries with the same CI growth rate are closer to each other within GVCs. In other words, there is a correlation between countries that have more bilateral TiVA. In spatial econometrics, this is referred to as spatial clustering. In this study, this space refers to GVC-based TiVA. A concept in spatial autocorrelation is biasedness, which occurs when spatial effects are not considered in estimating a regression.

Table 1. Spatial autocorrelation of CI growth

| | I | E(I) | SD(I) | Z | P-value |
|---|---|---|---|---|---|
| Moran's I | 0.05 | -0.01 | 0.02 | 2.71 | 0.00 |
| | C | E(C) | SD(C) | Z | P-value |
| Geary's c | 0.66 | 1.00 | 0.16 | -2.10 | 0.03 |

*2-tail test, null hypothesis: spatial randomization

Once the spatial autocorrelation is estimated, spatial data panel regressions are estimated to analyze the CI growth convergence of the countries. The results of SAR, SEM, and SDM are shown in Table 2. The results of fixed-effects (FE) panel regression are also reported to compare their coefficients with the spatial regressions.

Specification and model selection tests will be evaluated before interpreting the results. Based on the Wald test results, the significance of all regressions is confirmed. The Hausman specification test indicates that FE regression is preferred to random effects. Moreover, the model selection tests show that SDM is preferred to both SAR and SEM. Thus, it can be concluded that SDM is the best regression model.

The $\rho$ coefficient indicates that there is a positive spatial autocorrelation in regression (the $\lambda$ coefficient in SEM shows the same result). Considering Moran's I and Geary's C, since spatial regression involves spillover effects, the direct, indirect, and total effects (Table 3) should be calculated (using Equations 10 to 12) to interpret the coefficients of the variables. In all the regressions, the convergence rate ranges from 0 to 1, and conditional convergence is not rejected. Moreover, the explanatory power of the spatial regression is substantially higher and the conditional convergence rate is lower in FE regressions compared to spatial regressions. This can potentially be attributed to exclusion of spillover effects from the regression models. Therefore, it can be concluded that GVC-based spatial spillover effects increase the CI conditional convergence rate.

The direct negative effects of $Ln(CI)_{t-1}$ indicate that any period of high CI growth is followed by a period of low CI growth. In addition, its negative indirect effects indicate that there are GVC-based spillover effects on neighboring countries. As a result, the total effects of $Ln(CI)_{t-1}$ reveal the CI conditional convergence of the countries. Such effects are also observed on the real per capita income (Y), either directly or indirectly. Therefore, the increased real per capita income of countries (which is associated with economic growth and welfare) not only reduces their CI growth, but also decreases spillover effects on neighboring countries within GVCs. The results show that all the direct, indirect, and total effects of the urbanization rate (UR) are significantly positive. Since the increasing urban population is likely to increase fossil fuel consumption and $CO_2$ emissions more than GDP growth, the increased urbanization rate increases the CI growth of the countries both directly and indirectly by affecting their trading partners. Therefore, the urbanization rate can be the cause of the CI growth divergence of countries. The results also indicate that increased EI, regardless of per capita income and urbanization rate, is followed by increased CI growth. However, in cases where per capita income and urbanization rate are controlled, the EI coefficient is not statistically significant, and it could not be a determinant of CI growth and its conditional convergence.

The results demonstrate that all direct, indirect, and total effects of GVCs participation are significant. Thus, it can be stated that expanding participation in GVCs can help reduce CI growth and facilitate its conditional convergence. This correlation is more prominent when the indirect (spillover) effects of bilateral TiVA or GVC-based trade are taken into account. This means that expanding and boosting participation in GVCs will reduce the CI growth of the countries. In fact, expanding participation in GVCs not only reduces the CI growth of the countries, but also affects those of their trading partners. As noted earlier, expanding participation in GVCs can lead to technology advances, learning-by-doing, knowledge spillovers, economic growth, and higher welfare. These potential developments can help enhance CI by improving EI, encouraging the use of cleaner sources of energy, and increasing the demand for environmental protection laws and regulations.

Table 2. Estimated spatial panel data regressions

| Dependent Variable: $Ln(CI_t/CI_{t-1})$ | | Model | | | | | | | | | | | | | | | | |
|---|---|---|---|---|---|---|---|---|---|---|---|---|---|---|---|---|---|
| | | 1 | 2 | 3 | 4 | 5 | 6 | 7 | 8 | 9 | 10 | 11 | 12 | 13 | 14 | 15 | 16 |
| Method | | FE | SAR | SEM | SDM | FE | SAR | SEM | SDM | FE | SAR | SEM | SDM | FE | SAR | SEM | SDM |
| Independent Variable | | | | | | | | | | | | | | | | | |
| Main | $Ln(CI)_{t-1}$ | -.20a [.01] | -.20a [.01] | -.22a [.01] | -.21a [.01] | -.20a [.01] | -.20a [.01] | -.23a [.01] | -.22a [.01] | -.17a [.01] | -.17a [.01] | -.23a [.01] | -.21a [.01] | -.21a [.01] | -.22a [.01] | -.25a [.01] | -.23a [.01] |
| | $Ln(Y)_{t-1}$ | -.06a [.01] | -.06a [.01] | -.03b [.01] | -.03b [.01] | | | | | | | | | -.10a [.01] | -.1a [.02] | -.07a [.01] | -.06a [.02] |
| | $Ln(EI)_{t-1}$ | | | | | .05a [.01] | .05a [.01] | .03 [.01] | .02 [.01] | | | | | -.01 [.02] | -.01 [.02] | -.008 [.02] | -.004 [.02] |
| | $Ln(UR)_{t-1}$ | | | | | | | | | .01 [.04] | .01 [.04] | .20a [.05] | .15a [.05] | .17a [.05] | .17a [.05] | .26a [.05] | .23a [.05] |
| | $Ln(GVC)_{t-1}$ | -.03c [.02] | -.03c [.01] | -.007 [.01] | -.02 [.01] | -.05a [.01] | -.05a [.01] | -.009 [.02] | -.02 [.02] | -.06a [.02] | -.06a [.01] | -.01 [.02] | -.03c [.01] | -.05b [.02] | -.04b [.01] | -.017 [.02] | -.03c [.01] |
| Spatial | $\rho$ | | .40a [.07] | | .38a [.07] | | .40a [.07] | | .38a [.07] | | .40a [.07] | | .37a [.07] | | .40a [.07] | | .37a [.07] |
| | $\lambda$ | | | .47a [.06] | | | | .51a [.06] | | | | .59a [.05] | | | | .52a [.06] | |
| | $WLn(GVC)_{t-1}$ | | | | -.16a [.04] | | | | -.20a [.04] | | | | -.28a [.04] | | | | -.21a [.04] |
| Wald test statistic (p-value) | | 62 (.00) | 231 (.00) | 253 (.00) | 244 (.00) | 58 (.00) | 217 (.00) | 217 (.00) | 242 (.00) | 55 (.00) | 208 (.00) | 291 (.00) | 251 (.00) | 39 (.00) | 243 (.00) | 292 (.00) | 265 (.00) |
| Pseudo $R^2$ | | 0.16 | 0.37 | 0.38 | 0.38 | 0.10 | 0.35 | 0.37 | 0.38 | 0.08 | 0.34 | 0.35 | 0.38 | 0.15 | 0.36 | 0.38 | 0.39 |
| Model Selection Tests | Hausman test | chi2(3) =166.1 | Prob > chi2 = 0.00 | | | chi2(3) =158.4 | Prob > chi2 = 0.00 | | | chi2(4) =151.5 | Prob > chi2 = 0.00 | | | chi2(4) =176.6 | Prob > chi2 = 0.00 | | |
| | SAR vs SDM | chi2(1) = 11.8 | Prob > chi2 = 0.00 | | | chi2(1) =23.1 | Prob > chi2 = 0.00 | | | chi2(1) =38.6 | Prob > chi2 = 0.00 | | | chi2(1) =19.6 | Prob > chi2 = 0.00 | | |
| | SEM vs SDM | chi2(1) = 13.6 | Prob > chi2 = 0.00 | | | chi2(1) =27.9 | Prob > chi2 = 0.00 | | | chi2(1) =44.7 | Prob > chi2 = 0.00 | | | chi2(1) =22.8 | Prob > chi2 = 0.00 | | |

*a: p-value <0.01, b: p-value <0.05, c: p-value <0.1 **Standard Errors are in the brackets.

Table 3. Estimated direct, indirect, and total effects of the regressors and the conditional convergence rate of CI

| Dependent Variable: $Ln(CI_t/CI_{t-1})$ | | Model | | | | | | | | | | | | | | | |
|---|---|---|---|---|---|---|---|---|---|---|---|---|---|---|---|---|---|
| | | 1 | 2 | 3 | 4 | 5 | 6 | 7 | 8 | 9 | 10 | 11 | 12 | 13 | 14 | 15 | 16 |
| Method | | FE | SAR | SEM | SDM | FE | SAR | SEM | SDM | FE | SAR | SEM | SDM | FE | SAR | SEM | SDM |
| Independent Variable | | | | | | | | | | | | | | | | | |
| Direct Effect | $Ln(CI)_{t-1}$ | -.20[a] [.01] | -.21[a] [.01] | -.22[a] [.01] | -.21[a] [.01] | -.20[a] [.01] | -.20[a] [.01] | -.23[a] [.01] | -.22[a] [.01] | -.17[a] [.01] | -.17[a] [.01] | -.23[a] [.01] | -.21[a] [.01] | -.21[a] [.01] | -.22[a] [.01] | -.25[a] [.01] | -.23[a] [.01] |
| | $Ln(Y)_{t-1}$ | -.06[a] [.01] | -.06[a] [.01] | -.03[b] [.01] | -.03[b] [.01] | | | | | | | | | -.10[a] [.01] | -.10[a] [.02] | -.07[a] [.01] | -.06[a] [.03] |
| | $Ln(EI)_{t-1}$ | | | | | .05[a] [.01] | .05[a] [.01] | .03 [.01] | .02 [.01] | | | | | -.01 [.02] | -.01 [.02] | -.008 [.02] | -.003 [.02] |
| | $Ln(UR)_{t-1}$ | | | | | | | | | .01 [.04] | .01 [.04] | .20[a] [.05] | .15[a] [.05] | .17[a] [.05] | .17[a] [.05] | .26[a] [.05] | .23[a] [.05] |
| | $Ln(GVC)_{t-1}$ | -.03[c] [.02] | -.03[c] [.01] | -.007 [.01] | -.02 [.01] | -.05[a] [.01] | -.05[c] [.01] | -.009 [.02] | -.02 [.01] | -.06[a] [.02] | -.06[a] [.01] | -.01 [.02] | -.03[c] [.01] | -.05[b] [.02] | -.04[a] [.01] | -.017 [.02] | -.03[c] [.01] |
| Indirect Effect | $Ln(CI)_{t-1}$ | | -.14[a] [.04] | | -.13[a] [.05] | | -.14[a] [.04] | | -.14[a] [.04] | | -.12[a] [.04] | | -.13[a] [.04] | | -.15[a] [.05] | | -.15[a] [.05] |
| | $Ln(Y)_{t-1}$ | | -.04[b] [.01] | | -.02[c] [.012] | | | | .51[a] [.06] | | | | | | -.07[b] [.02] | | -.04[a] [.01] |
| | $Ln(EI)_{t-1}$ | | | | | | .03[b] [.01] | | .01 [.01] | | | | | | -.007 [.01] | | -.002 [.01] |
| | $Ln(UR)_{t-1}$ | | | | | | | | | | .007 [.03] | | .09[b] [.04] | | .12[b] [.05] | | .14[b] [.06] |
| | $Ln(GVC)_{t-1}$ | | -.02 [.01] | | -.28[a] [.08] | | -.03[b] [.01] | | -.3[b] [.07] | | -.04[b] [.01] | | -.47[a] [.09] | | -.03[c] [.01] | | -.36[a] [.09] |
| Total Effect | $Ln(CI)_{t-1}$ | -.20[a] [.01] | -.35[a] [.05] | -.22[a] [.01] | -.35[a] [.05] | -.20[a] [.01] | -.35[a] [.05] | -.23[a] [.01] | -.36[a] [.05] | -.17[a] [.01] | -.35[a] [.05] | -.23[a] [.01] | -.34[a] [.04] | -.21[a] [.01] | -.37[a] [.06] | -.25[a] [.01] | -.39[a] [.06] |
| | $Ln(Y)_{t-1}$ | -.06[a] [.01] | -.11[a] [.03] | -.03[b] [.01] | -.05[b] [.02] | | | | | | | | | -.10[a] [.01] | -.17[a] [.04] | -.07[a] [.01] | -.10[a] [.03] |
| | $Ln(EI)_{t-1}$ | | | | | .05[a] [.01] | .08[b] [.03] | .03 [.01] | .04 [.03] | | | | | -.01 [.02] | -.01 [.03] | -.008 [.02] | -.006 [.03] |
| | $Ln(UR)_{t-1}$ | | | | | | | | | .01 [.04] | .01 [.07] | .20[a] [.05] | .25[a] [.08] | .17[a] [.05] | .29[a] [.09] | .26[a] [.05] | .37[a] [.09] |
| | $Ln(GVC)_{t-1}$ | -.03[c] [.02] | -.05[c] [.03] | -.007 [.01] | -.30[a] [.08] | -.05[a] [.01] | -.09[a] [.03] | -.009 [.02] | -.36[a] [.07] | -.06[a] [.02] | -.10[a] [.03] | -.01 [.02] | -.51[a] [.09] | -.05[b] [.02] | -.08[c] [.03] | -.017 [.02] | -.40[a] [.09] |
| Convergence rate | | 0.22 | 0.43 | 0.24 | 0.43 | 0.22 | 0.43 | 0.26 | 0.44 | 0.18 | 0.43 | 0.26 | 0.41 | 0.23 | 0.46 | 0.28 | 0.49 |

*a: p-value <0.01, b: p-value <0.05, c: p-value <0.1 **Standard Errors are in the brackets.

# 6. Conclusion

Having transformed the structure and processes of production in most developing and developed countries, GVCs provide a new form of trade that allows countries to benefit from their comparative advantages in production tasks. Apart from the economic benefits of GVC-based trade and expanding participation in GVCs, their environmental advantages and disadvantages are also among controversial topics in the economic literature.

This study analyzed the role of GVCs in the environmental performance of countries and sought to determine whether GVC-based trade could facilitate the CI convergence of the countries. The study also evaluated the effects of participation in GVCs on the CI growth of the countries. Spatial statistics were employed to model and analyze the data collected from 101 countries in the period from 1997 to 2014. Moreover, the economic distance between the countries was defined in terms of their bilateral TiVA within GVCs. The results of statistical tests showed that there was a CI growth correlation between countries that had more bilateral GVC-based trade. In addition, considering the GVC-based trade between countries, the conditional convergence of countries in terms of the CI growth was confirmed. The results also demonstrate that expanding participation in GVCs not only helps reduce the CI growth of the countries, but it also affects the CI growth of their trading partners (through spillover effects). Hence, the environmental performance of a country with GVCs-based trade partners that have better environmental performance will relatively improve.

It can also be stated that this new form of trade provides various capacities, especially for developing countries, to benefit from the knowledge and technology of their trading partners for economic and environmental purposes. In this regard, the necessity of adapting local production processes to global processes within GVCs can encourage the promotion of production processes and their adaptation to the global production structure. Additionally, the environmental policies of countries within the context of GVC-based trade are more likely to become more convergent. For instance, multinational enterprises that outsource their production processes within GVCs will probably not only transfer technology and knowledge to their partners in other countries, but also require them to comply with certain laws and regulations, including environmental standards. Furthermore, GVCs can serve as a potential platform for improving the environmental performance of countries (at least reduce their CI growth) and converging their environmental indicators.

According to the findings, the capacities of GVCs should be considered in environmental policies and measures to combat climate change. The results showed that expanding participation in GVCs improved the carbon efficiency of the countries, *i.e.*, an important and fundamental (albeit insufficient) step in advancing environmental goals, even with the existence of carbon leakage in trade. Moreover, increasing TiVA with countries on the path of improving carbon efficiency can be an appropriate strategy for developing countries to enhance their environmental performance.

This study contributes to the empirical literature on the link between trade and the environment by analyzing the CI correlation and conditional convergence of countries within the context of GVC-based trade. Nevertheless, more studies must be conducted on the subject, especially on the emission of greenhouse gases and pollutants such as $SO_2$ to gain a better insight into the role of GVCs in the convergence of the countries' environmental performance. The scale, composition, and technical effects of expanding GVCs on the environmental performance of countries can also be an important area of research for future studies to separately analyze different mechanisms of impacts for the GVCs on the environment.

# References

Anselin, L., 2003a. Spatial externalities, spatial multipliers, and spatial econometrics. Int. Reg. Sci. Rev. 26, 153–166. https://doi.org/10.1177/0160017602250972

Anselin, L., 2003b. Spatial Econometrics, in: A Companion to Theoretical Econometrics. John Wiley & Sons, Ltd, pp. 310–330. https://doi.org/https://doi.org/10.1002/9780470996249.ch15

Antweiler, W., Copeland, B.R., Taylor, M.S., 2001. Is free trade good for the environment? Am. Econ. Rev. 91, 877–908. https://doi.org/10.1257/aer.91.4.877

Apergis, N., Payne, J.E., 2020. NAFTA and the convergence of $CO_2$ emissions intensity and its determinants. Int. Econ. 161, 1–9. https://doi.org/10.1016/j.inteco.2019.10.002

Aslam, A., Novta, N., Rodrigues-Bastos, F., 2017. Calculating Trade in Value Added, WP/17/178, July 2017.

Awad, A., 2019. Does economic integration damage or benefit the environment? Africa's experience. Energy Policy 132, 991–999. https://doi.org/10.1016/j.enpol.2019.06.072

Baghdadi, L., Martinez-Zarzoso, I., Zitouna, H., 2013. Are RTA agreements with environmental provisions reducing emissions? J. Int. Econ. 90, 378–390. https://doi.org/10.1016/j.jinteco.2013.04.001

Baldwin, R., 2017. The great convergence: Information technology and the New Globalization, Ekonomicheskaya Sotsiologiya. https://doi.org/10.17323/1726-3247-2017-5-40-51

Bernhardt, T., Pollak, R., 2016. Economic and social upgrading dynamics in global manufacturing value chains: A comparative analysis. Environ. Plan. A 48, 1220–1243. https://doi.org/10.1177/0308518X15614683

Casella, B., Bolwijn, R., Moran, D., Kanemoto, K., 2019. UNCTAD insights: Improving the analysis of global value chains: the UNCTAD-Eora Database. Transnatl. Corp. 26, 115–142. https://doi.org/10.18356/3aad0f6a-en

Cherniwchan, J., Copeland, B.R., Taylor, M.S., 2017. Trade and the environment: New methods, measurements, and results. Annu. Rev. Econom. 9, 59–85. https://doi.org/10.1146/annurev-economics-063016-103756

Conley, T.G., Topa, G., 2002. Socio-economic distance and spatial patterns in unemployment. J. Appl. Econom. 17, 303–327.

Copeland, B.R., Taylor, M.S., 2013. Trade and the environment: Theory and evidence, Trade and the Environment: Theory and Evidence. https://doi.org/10.2307/3552527

Dorrucci, E., Gunnella, V., Al-Haschimi, A., Benkovskis, K., Chiacchio, F., de Soyres, F., Lupidio, B. Di, Fidora, M., Franco-Bedoya, S., Frohm, E., Gradeva, K., López-Garcia, P., Koester, G., Nickel, C., Osbat, C., Pavlova, E., Schmitz, M., Schroth, J., Skudelny, F., Tagliabracci, A., Vaccarino, E., Wörz, J., 2019. The impact of global value chains on the euro area economy. European Central Bank (ECB), Frankfurt a. M. https://doi.org/10.2866/870210

Frankel, J.A., Rose, A.K., 2005. Is trade good or bad for the environment? sorting out the causality. Rev. Econ. Stat. 87, 85–91. https://doi.org/10.1162/0034653053327577

Gangodagamage, C., Zhou, X., Lin, H., 2008. Autocorrelation, Spatial, in:

Shekhar, S., Xiong, H. (Eds.), Encyclopedia of GIS. Springer US, Boston, MA, pp. 32–37. https://doi.org/10.1007/978-0-387-35973-1_83

Gerlagh, R., Kuik, O., 2014. Spill or leak? Carbon leakage with international technology spillovers: A CGE analysis. Energy Econ. 45, 381–388. https://doi.org/10.1016/j.eneco.2014.07.017

Han, L., Han, B., Shi, X., Su, B., Lv, X., Lei, X., 2018. Energy efficiency convergence across countries in the context of China's Belt and Road initiative. Appl. Energy 213, 112–122. https://doi.org/10.1016/j.apenergy.2018.01.030

Hanson, G.H., 2012. The rise of middle kingdoms: Emerging economies in global trade. J. Econ. Perspect. 26, 41–64. https://doi.org/10.1257/jep.26.2.41

Holzinger, K., Knill, C., Sommerer, T., 2008. Environmental policy convergence: The impact of international harmonization, transnational communication, and regulatory competition. Int. Organ. 62, 553–587. https://doi.org/10.1017/S002081830808020X

Huang, J., Liu, C., Chen, S., Huang, X., Hao, Y., 2019. The convergence characteristics of China's carbon intensity: Evidence from a dynamic spatial panel approach. Sci. Total Environ. 668, 685–695. https://doi.org/10.1016/j.scitotenv.2019.02.413

Huang, R., Chen, G., Lv, G., Malik, A., Shi, X., Xie, X., 2020. The effect of technology spillover on CO2 emissions embodied in China-Australia trade. Energy Policy 144, 111544. https://doi.org/10.1016/j.enpol.2020.111544

Huang, Z., Zhang, H., Duan, H., 2019. Nonlinear globalization threshold effect of energy intensity convergence in Belt and Road countries. J. Clean. Prod. 237. https://doi.org/10.1016/j.jclepro.2019.117750

Ignatenko, A., Raei, F., Mircheva, B., 2019. Global Value Chains: What are the Benefits and Why Do Countries Participate? IMF Working Paper European Department Global Value Chains: What are the Benefits and Why Do Countries Participate ? IMF Work. Pap.

Jaffe, A.B., Newell, R.G., Stavins, R.N., 2002. Environmental policy and technological change. Environ. Resour. Econ. 22, 41–70.

Jangam, B.P., Rath, B.N., 2020. Cross-country convergence in global value chains: Evidence from club convergence analysis. Int. Econ. 163, 134–146. https://doi.org/10.1016/j.inteco.2020.06.002

Jiang, L., Folmer, H., Ji, M., Zhou, P., 2018. Revisiting cross-province energy intensity convergence in China: A spatial panel analysis. Energy Policy 121, 252–263. https://doi.org/10.1016/j.enpol.2018.06.043

Jiang, M., An, H., Gao, X., Liu, S., Xi, X., 2019. Factors driving global carbon emissions: A complex network perspective. Resour. Conserv. Recycl. 146, 431–440. https://doi.org/10.1016/j.resconrec.2019.04.012

Kalkhan, M.A., 2011. Spatial statistics: geospatial information modeling and thematic mapping. CRC press.

Kaltenegger, O., Löschel, A., Pothen, F., 2017. The effect of globalisation on energy footprints: Disentangling the links of global value chains. Energy Econ. 68, 148–168. https://doi.org/10.1016/j.eneco.2018.01.008

Keller, W., 2010. International trade, foreign direct investment, and technology spillovers, 1st ed, Handbook of the Economics of Innovation. Elsevier B.V. https://doi.org/10.1016/S0169-7218(10)02003-4

Koopman, R., Wang, Z., Wei, S.J., 2014. Tracing value-added and double counting in gross exports. Am. Econ. Rev. 104, 459–494.

https://doi.org/10.1257/aer.104.2.459

Kummritz, V., 2016. Do Global Value Chains Cause Industrial Development?, CTEI Working Paper No 2016-01.

LeSage, J., Pace, R.K., 2009. Introduction to Spatial Econometrics. Chapman and Hall/CRC. https://doi.org/10.1201/9781420064254

LeSage, J.P., Fischer, M.M., 2008. Spatial growth regressions: Model specification, estimation and interpretation. Spat. Econ. Anal. 3, 275–304. https://doi.org/10.1080/17421770802353758

Levin, A., Lin, C.F., Chu, C.S.J., 2002. Unit root tests in panel data: Asymptotic and finite-sample properties. J. Econom. 108, 1–24. https://doi.org/10.1016/S0304-4076(01)00098-7

Liu, H., Li, J., Long, H., Li, Z., Le, C., 2018. Promoting energy and environmental efficiency within a positive feedback loop: Insights from global value chain. Energy Policy 121, 175–184. https://doi.org/10.1016/j.enpol.2018.06.024

López, L.A., Cadarso, M.Á., Zafrilla, J., Arce, G., 2019. The carbon footprint of the U.S. multinationals' foreign affiliates. Nat. Commun. 10, 1–11. https://doi.org/10.1038/s41467-019-09473-7

Lovely, M., Popp, D., 2011. Trade, technology, and the environment: Does access to technology promote environmental regulation? J. Environ. Econ. Manage. 61, 16–35. https://doi.org/10.1016/j.jeem.2010.08.003

Markusen, J.R., 1984. Multinationals, multi-plant economies, and the gains from trade. J. Int. Econ. 16, 205–226.

Nemati, M., Hu, W., Reed, M., 2019. Are free trade agreements good for the environment? A panel data analysis. Rev. Dev. Econ. 23, 435–453.

https://doi.org/10.1111/rode.12554

Nicoletti, G., Golub, S.S., Hajkova, D., Mirza, D., Yoo, K.-Y., 2003. Policies and international integration: influences on trade and foreign direct investment (No. 359), OECD Economics Department Working Papers. https://doi.org/http://dx.doi.org/10.1787/062321126487

Piermartini, R., Rubínová, S., 2014. Knowledge spillovers through international supply chains, WTO Staff Working Papers.

Pigato, M., Black, S.J., Dussaux, D., Mao, Z., McKenna, M., Rafaty, R., Touboul, S., 2020. Technology Transfer and Innovation for Low-Carbon Development. Washington, DC: World Bank. https://doi.org/10.1596/978-1-4648-1500-3

Qi, S.Z., Peng, H.R., Zhang, Y.J., 2019. Energy intensity convergence in Belt and Road Initiative (BRI) countries: What role does China-BRI trade play? J. Clean. Prod. 239, 118022. https://doi.org/10.1016/j.jclepro.2019.118022

Rodrik, D., 2018. New Technologies, Global Value Chains, and Developing Economies. Oxford. United Kingdom, Cambridge, MA. https://doi.org/10.3386/w25164

Servén, L., Abate, G.D., 2020. Adding space to the international business cycle. J. Macroecon. 65. https://doi.org/10.1016/j.jmacro.2020.103211

Shen, J., 2008. Trade liberalization and environmental degradation in China. Appl. Econ. 40, 997–1004. https://doi.org/10.1080/00036840600771148

Sun, C., Li, Z., Ma, T., He, R., 2019. Carbon efficiency and international specialization position: Evidence from global value chain position index of manufacture. Energy Policy 128, 235–242. https://doi.org/10.1016/j.enpol.2018.12.058

Taglioni, D., Winkler, D., Taglioni, D., Winkler, D., 2016. Making Global Value Chains Work for Development, in: Making Global Value Chains Work for Development. The World Bank, pp. i–xxii. https://doi.org/10.1596/978-1-4648-0157-0_fm

Thomakos, D.D., Alexopoulos, T.A., 2016. Carbon intensity as a proxy for environmental performance and the informational content of the EPI. Energy Policy 94, 179–190. https://doi.org/10.1016/j.enpol.2016.03.030

Wan, J., Baylis, K., Mulder, P., 2015. Trade-facilitated technology spillovers in energy productivity convergence processes across EU countries. Energy Econ. 48, 253–264. https://doi.org/10.1016/j.eneco.2014.12.014

Wang, Z., Wei, S.-J., Yu, X., Zhu, K., 2017. NBER WORKING PAPER SERIES MEASURES OF PARTICIPATION IN GLOBAL VALUE CHAINS AND GLOBAL BUSINESS CYCLES Measures of Participation in Global Value Chains and Global Business Cycles. NBER Work. Pap.

WDR, 2020. World Development Report 2020: Trading for Development in the Age of Global Value Chains. Washington, DC: World Bank, Washington, DC. https://doi.org/10.1596/978-1-4648-1457-0

Weber, S., Gerlagh, R., Mathys, N.A., Moran, D., 2021. CO2 embodied in trade: trends and fossil fuel drivers. Environ. Sci. Pollut. Res. 1–19. https://doi.org/10.1007/s11356-020-12178-w

Xin-gang, Z., Yuan-feng, Z., Yan-bin, L., 2019. The spillovers of foreign direct investment and the convergence of energy intensity. J. Clean. Prod. 206, 611–621. https://doi.org/10.1016/j.jclepro.2018.09.225

Yasmeen, R., Li, Y., Hafeez, M., 2019. Tracing the trade–pollution nexus in global value chains: evidence from air pollution indicators. Environ. Sci. Pollut. Res. 26, 5221–5233. https://doi.org/10.1007/s11356-018-3956-0

You, W., Lv, Z., 2018. Spillover effects of economic globalization on CO2 emissions: A spatial panel approach. Energy Econ. 73, 248–257. https://doi.org/10.1016/j.eneco.2018.05.016

# Appendix A. List of Countries

## Table A1. List of countries in sample

| | | |
|---|---|---|
| Albania | Hungary | Romania |
| Argentina | Iceland | Russian Federation |
| Australia | India | Saudi Arabia |
| Austria | Indonesia | Senegal |
| Azerbaijan | Iran, Islamic Rep. | Singapore |
| Bangladesh | Iraq | Slovak Republic |
| Belgium | Ireland | Slovenia |
| Bolivia | Israel | South Africa |
| Bosnia and Herzegovina | Italy | Spain |
| Botswana | Jamaica | Sri Lanka |
| Brazil | Japan | Sweden |
| Bulgaria | Jordan | Switzerland |
| Cambodia | Kazakhstan | Tajikistan |
| Cameroon | Korea, Rep. | Thailand |
| Canada | Kyrgyz Republic | Togo |
| Chile | Latvia | Tunisia |
| China | Lebanon | Turkey |
| Colombia | Lithuania | Ukraine |
| Costa Rica | Luxembourg | United Kingdom |
| Cote d'Ivoire | Malaysia | United States |
| Croatia | Malta | Uruguay |
| Cuba | Mauritius | |
| Cyprus | Mexico | |
| Czech Republic | Mongolia | |
| Denmark | Morocco | |
| Dominican Republic | Mozambique | |
| Ecuador | Namibia | |
| Egypt, Arab Rep. | Nepal | |
| El Salvador | Netherlands | |
| Estonia | New Zealand | |
| Finland | Nicaragua | |
| France | Nigeria | |
| Gabon | Norway | |
| Georgia | Pakistan | |
| Germany | Panama | |
| Ghana | Paraguay | |
| Greece | Peru | |
| Guatemala | Philippines | |
| Haiti | Poland | |
| Honduras | Portugal | |

# Appendix B. Unit root tests

**Table B1. Levin, Lin and Chu (2002) unit root tests**

| Variable | Test Statistic (Adjusted t*) | Probability |
|---|---|---|
| $Ln(CI)$ | -3.13 | 0.0009 |
| $Ln(Y)$ | -3.51 | 0.0002 |
| $Ln(EI)$ | -4.13 | 0.0000 |
| $Ln(UR)$ | -4.23 | 0.0000 |
| $Ln(GVC)$ | -4.82 | 0.0000 |

Null Hypothesis: Unit root (common unit root process)